# Phototgyrotactic concentration of a population of swimming microalgae across a porous layer

Praneet Prakash[1], Ottavio A. Croze [2,*]

[1] Department of Applied Mathematics and Theoretical Physics, Centre for Mathematical Sciences, University of Cambridge, Cambridge CB3 0WA, United Kingdom
[2] School of Mathematics, Statistics and Physics, Newcastle University, Newcastle upon Tyne NE1 7RU, United Kingdom

*Corresponding Author
otti.croze@newcastle.ac.uk

Keywords: Microswimmers, Microalgae, Phototaxis, Light environment, Porous media.

Abstract

The light environment controls the swimming of microalgae through a light-seeking and avoiding behaviour, which is known as phototaxis. In this work, we exploit phototaxis to control the migration and concentration of populations of the soil microalga *Chlamydomonas reinhardtii*. By imaging a suspension of these microalgae in a cuvette illuminated from above by blue light, we study how phototaxis changes the stability of the suspension and demonstrate how a thin, porous layer at the top of the cuvette prevents phototaxing microalgae from sinking, leading to the up-concentration of the microalgae in the region above the porous layer. We discuss the potential implications of our findings for microalgae in biotechnological applications and the natural environment.

## 1    Introduction

Environmental stimuli, such as chemical gradients, gravity, light and flow shear, bias the motion of swimming microorganisms [1–3]. At the level of a population, these biases cause the formation of spectacular, often macroscopic, patterns. Inasmuch as they cause cells to congregate and interact, these patterns can be considered a form of social behaviour. Paradigmatic examples of pattern formation in swimming microbes are the waves exhibited by bacteria [3] and slime mold [6] sensing chemical gradients (chemotaxis), or the bioconvection patterns formed by ciliates and microalgae [7], responding to a combination of gravity and flow shear (gyrotaxis). More specifically, the latter is a bias resulting from the combination of a torque on a swimmer due to shear in the flow and one due to gravity, caused by asymmetry in body shape, mass distribution and/or between body and flagella [8].

Recent decades have seen a marked increase in the mechanistic understanding of how biases act at the individual swimmer level and how this affects macroscopic patterns. For example, mathematical models of flowing and dispersing gyrotactic suspensions of microalgae [10–13] have been compared with measurements of algae in an uniformly rotating flow [13], sheared bioconvection patterns [14], populations of microalgae dispersing in pipe flow [15], and laboratory versions of oceanic thin layers [16]. For comprehensive summaries of current work, we refer the reader to reviews covering recent progress in the physics of swimming microbes [17,18] and bioconvection [7]. This area of research is also closely related to active matter [19] comprising biological swimmers, as we have just described, but also synthetic [20] and biohybrid ones [21–23].

In this study, we focus on how light can be used to control and concentrate a suspension of *Chlamydomonas reinhardtii* microalgae. The bias of swimming by light is known as phototaxis and is an adaptation that allows *C. reinhardtii* and other photosynthetic microorganisms to find optimal levels of light needed to grow [24,25]. Recent experimental and theoretical studies have demonstrated how exposing a suspension of microalgae to light can dramatically alter the patterns they form, and even generate new ones [21]. Bees and Williams investigated how white light from above and below a suspension of the microalga *Chlamydomonas augustae* in a Petri dish alters the stability of bioconvection patterns, quantified by measuring the dominant initial pattern wavelength [9]. In the absence of this phototactic stimulation combined with gyrotaxis (photogyrotaxis), bioconvection patterns form in shallow layer suspensions, e.g. a thin layer of fluid in a Petri dish, as a result of the tendency of microalgae to swim upwards (gravitaxis) and form a dense layer of cells (denser than the fluid they are suspended in) at the top of the suspension [7]. This is unstable and results in sinking 'plumes', which drive a convective pattern, reinforced by response to the flow (gyrotaxis), which enhances the instability by driving cells towards downwelling plumes. As well as photogyrotactic changes to existing bioconvection patterns, recent investigations have also explored how shining light into a suspension can stimulate patterns that would not otherwise be there. For example, bioconvection patterns for the microalga *Euglena gracilis* were induced by illuminating a Hele-Shaw cell from below [27]. In the absence of light, the patterns vanished. More recently, a green laser light was observed by Dervaux *et al.* to induce bioconvective instabilities in a shallow suspension of *C. reinhardtii*, which was described by a simplified model of photogyrotaxis [28]. A study by Arrieta *et al.* also demonstrated how quickly bioconvective structures can be created, and even reconfigured by light, using it to generate 'blinking plumes'; the study also provided a model of this (ignoring gyrotactic effects), and reported good agreement with the experimental observations [26].

Aside from some of the studies above, several investigations in the literature have provided theoretical analyses of bioconvection in the presence of phototaxis. These have recently been reviewed comprehensively [7,21]. We will discuss briefly here only the model by Williams and Bees [9], which includes phototactic and gyrotactic effects, and encompasses several simpler models that have been recently proposed. The model equations, summarized in Appendix A, describe the coupled dynamics of fluid flow, described by a Navier-Stokes equation, and a population of swimmers, described by a continuity equation. The probability density function (PDF) for the swimmer orientation obeys a Fokker-Planck equation, with a deterministic bias due to the combined action of flow, gravity and light. Taking moments of this PDF provides the mean swimming velocity and diffusivity in the continuity equation. Williams and Bees considered three alternative models to describe the effect of phototaxis on the swimmers [9]. In model A, the speed of the cells is dependent on light intensity (photokinesis), while gravitaxis and gyrotaxis are not affected. In model B, light causes a change in the bottom-heaviness of the cells, inducing an effective gravi/gyrotactic torque. In model C, cells respond directly to an effective torque due to light, dependent either of the light direction or the gradient of its intensity (the latter was also used by [24,26]). Williams and Bees used their model to predict the stability of bioconvection patterns for a suspension illuminated from above and below, in qualitative agreement with the experiments with *C. augustae* microalgae in a Petri dish mentioned above [30].

Thus, it is well established that light perturbs, and drives instabilities in, suspensions of phototactic microalgae, visibly causing the concentration of cells. However, the systematic concentration of microalgae at a given location exploiting photogyrotaxis, which was suggested by Kessler as early as 1982 [31], has hitherto not been demonstrated. In this study, we report the first 'milliliter-scale' experiments to show how photogyrotactic microalgae can be concentrated above a porous layer of beads overlaid onto a metal mesh. We show that it is the unique combination of phototaxis and porous media that permits the optimal concentration of microalgae. We also observe interesting photogyrotactic accumulations in the suspension, which have not been previously reported. An 'essential' model to account for the temporal evolution of the average concentration of cells above the porous layer and for their initial spatial distribution is also developed, leaving a full theoretical analysis of the photogyrotactic dynamics leading to this concentration for future work. Finally, we discuss how, a scaled-up version of our set-up could provide the basis for a new and efficient method to harvest swimming microalgae industrially. This is desirable since harvesting microalgae industrially is expensive (up to 20-30% of the total production costs [32]), and represents a bottleneck in the production of bioproducts from microalgae.

## 2  Materials and methods

### 2.1  Experimental methods

We used the wild-type algal strain *Chlamydomonas reinhardtii* (CC125) for our studies. Single colonies of these algae were picked from slant cultures and inoculated into Tris-minimal growth media (Supplemental Material (SM) Section 1). These media are based on the standard TAP medium [33], but omit acetic acid and HCl is used to titrate to pH 7. Liquid cultures of the microalgae were then grown in a 14:10 h light-dark cycle on a rotary shaker at 100 rpm and continuously bubbled with air, as in [34]. The shaking incubator (Infors Minitron) was maintained at a temperature of $25\,^{\circ}\mathrm{C}$, and provided photosynthetically active radiation (PAR) at $315 - 325\ \mu\mathrm{mol/m^2 s}$, as measured with a PAR meter (Skye SKP200). It took around $7 - 10$ days for a culture to reach a concentration of $1 - 2$ million cells/mL. Thereafter, it was sub-cultured by mixing 10 mL of grown algae into 140 mL of fresh Tris-minimal media until the cell count, measured with a Z2 Coulter counter (Beckman Coulter, Brea, CA), reached 1.5 million/mL; this took about a week. Subsequently, algae were diluted everyday by replacing 50 ml of the culture with fresh media. This protocol maintains the algal count between $1.2 - 1.5$ million/mL with mean diameter of $4.5 - 5\ \mu\mathrm{m}$ (estimated by using the Coulter counter); the subcultures can be used for $10 - 15$ days. Care was taken to do experiments with microalgae harvested during the light phase of the growing cycle to avoid variations in the swimming parameters, and in particular the swimming speed, which have been observed at the onset the dark phase [34]. All the experiments were carried out in square plastic cuvettes of external dimensions $12.5 \times 12.5 \times 45$ mm$^3$ (Sigma-Aldrich, filling volume 2.5 mL) filled with 2 ml of algal suspension. The imaging was performed using a monochrome CMOS camera (Pointgrey, Grasshopper3 GS3-U3-23S6M) fitted with a macro lens (Sigma 17-70mm f2.8-4). The cuvette was illuminated from the side by a red (660 nm) square $100 \times 100$ mm LED array (Advanced Illumination BL1960, Rochester, VT, USA), as shown in Figure 1a. This illumination was used as it allowed to image the suspension laterally without triggering a phototactic response [3]. The concentration of microalgae in the cuvette was estimated from the transmitted light intensity across the short dimension of the cuvette by applying the Lambert-Beer law: the intensity recorded by the camera (measured in arbitrary units, a.u.) can be converted into algal concentration (million/mL) from the calibration curve shown in SM Figure 1. The intensity decays as $I = I_o \exp\left(-A.C\right)$, where $I_o = 179$ is the intensity in arbitrary units in the presence of cuvette containing just Tris-min medium, $A = 0.22$ is the attenuation coefficient and $C$ is the algal concentration in million/mL. This exponential decay provides a mapping to concentration, with an excellent fit for intensity data higher than 50 a.u., and an R-squared value of 0.99 for a fit across the range of values (see SM Figure 1). In the experiments described below, the swimmer concentration was then quantified from images by first measuring integrated pixel intensity in selected regions (see e.g. SM Figure 2) of the cuvette using ImageJ, and then mapping to actual concentration values using the calibration curve just described.

For the phototaxis experiments, a blue LED (Thorlabs M470L2, nominal wavelength 470 nm) is mounted above the cuvette at a distance of 47 mm from its base. Using a PAR meter, the light intensity at the base of cuvette containing only the media was $16 - 18$ µmol/m²s, whereas the intensity immediately below the LED is $150 - 160$ µmol/m²s. The cuvette is separated into an upper 'harvest' and a lower 'reservoir' region by a porous layer of glass beads. The latter was achieved by folding a rectangular wire mesh so that it attaches to a cuvette, and overlaying it with glass beads of various weights, as shown in Figure 1b. To initialize experiments, first an empty cuvette was filled with an algal suspension approximately up to the mesh height and thereafter the mesh was installed. To make a porous layer of various thicknesses, beads of appropriate weight were placed over the mesh. Finally, more algal suspension was poured from the top to create a harvest region of height $\approx 0.5$ cm. The experiments reported below also considered the case of a bare mesh with no beads.

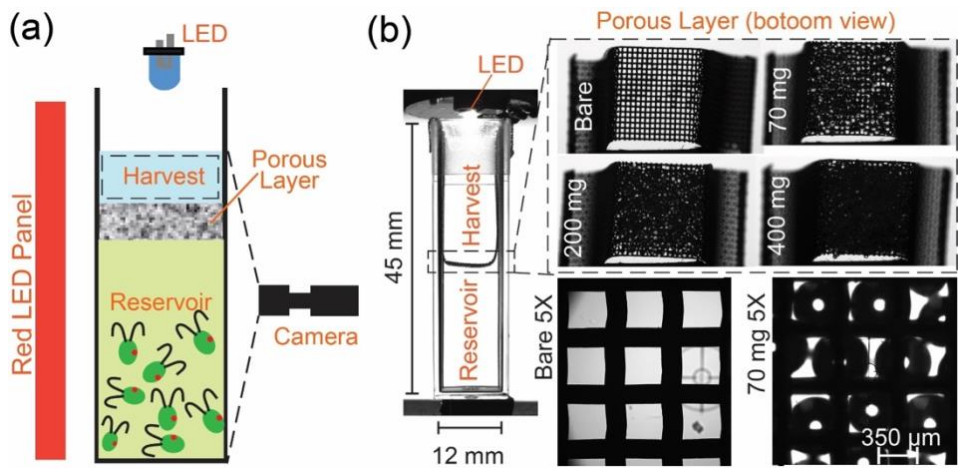

**Figure 1.** Schematic showing lower reservoir and upper harvest regions, separated by a wire mesh overlaid with beads: the porous layer. **(a)** A blue LED is mounted on the top of the cuvette to create a phototactic bias. For imaging the suspension and concentration calibration, a deep red LED illuminates the cuvette from the side (deep red light does not elicit phototaxis [3]). **(b)** Glass beads of diameter $425 - 600$ µm were used to create the porous layer. 70 mg of bead results in single layer, 200 mg – 2 to 3 layers, and 400 mg – 4 to 5 layers, respectively.

## 2.2 Essential model of concentration

We present here the details of a simplified model of the concentration of swimming microalgae into the upper 'harvest' region by light. The model describes the case of a suspension of microalgae with a porous layer near the top (mesh+beads), as shown in Figure 1, and we shall also apply it below to consider the case of a bare mesh. For the mesh+beads case, the suspension of microalgae is divided into three regions, an upper harvest region (u), a porous layer region (p) and a lower 'reservoir' region (l). Photogyrotactic migration delivers microalgae to the upper region from the lower region through the porous region. As evident from our results and discussion (see section 3.4 and 4 below), the dynamics underpinning the concentration are complex; the challenge of describing them with a full photogyrotactic model is beyond the scope of this paper. Instead, we seek here to formulate a model to capture the essential features of the concentration process into the upper harvest region. We make the reasonable simplifying assumption that: i) the average concentrations in the upper, porous and lower regions evolve slowly compared to the observed photogyrotactic dynamics; we consider here spatial and temporal variations separately, and assume a steady state for the fast dynamics in the upper region. We further assume that ii) phototactic migration is the dominant process and brings cells to from the lower region to the porous region, with swimmers migrating straight upwards toward the light (there is no dependence on light gradients, only light direction) at the maximum phototactic speed, equal to the mean swimming speed of the population, $V_s$. In the lower region, we assume that iii) the mean concentration is representative of the concentration of cells swimming into the porous layer. In the porous layer region, we assume that: iv) the speed of the swimmers is slowed down by collisions with the porous medium, but the swimming direction continues on average to be upwardly directed by phototaxis. In the upper harvest region, as well as the average concentration dynamics, we also consider a $0^{th}$ order spatial model of phototactic concentration. To set this up, as assumed above, we posit that there is a separation of timescales between the migration of cells from the lower region (slow) and the redistribution of cells in the upper region (fast). We further assume that: v) the effect of flow is negligible prior to the formation of the plume from the upper surface (see Figure 4b); vi) upward phototactic swimming at the maximum speed $V_s$ and diffusion dominate the fast suspension dynamics (gyrotactic effects are negligible); vii) diffusion is assumed approximately isotropic; viii) the meniscus at the top of the suspension is flat (any effects of curvature are neglected).

With the assumptions above, denoting by $\bar{c}_i$ the average concentrations in regions $i = u, p, l$ for the upper, porous and lower regions, respectively, and with A the cross-sectional area of the cuvette, the average cell numbers $\bar{N}_i$ in the three regions evolve according to the following balance equations:

$$\frac{d\bar{N}_l}{dt} = - A \, V_s \, \bar{c}_l, \tag{1a}$$

$$\frac{d\bar{N}_p}{dt} = A \, V_s \, \bar{c}_l - A \, V_{eff} \, \bar{c}_p, \tag{1b}$$

$$\frac{d\bar{N}_u}{dt} = A \, V_{eff} \, \bar{c}_p. \tag{1c}$$

Equation (1a) describes the loss of cells from the lower region due to the phototactic flux of cells, of concentration $\bar{c}_l$, swimming into the porous region at speed $V_s$. Correspondingly, the porous region, as described by equation (1b), gains an equal and opposite flux. This region also has a loss term due to cells, of concentration $\bar{c}_p$, swimming at a speed $V_{eff}$ into the upper region. The upper harvest region, as shown in equation (1c), has an equal and opposite gain. The speed $V_{eff}$ is the effective swimming speed of the microalgae within the porous layer, which is given by [35]

$$V_{eff} = V_s \frac{\tau_c}{\tau} + \frac{l_{obs}}{\tau}, \tag{2}$$

where $V_s$ is the 'free' mean swimming speed of the microalgae, $\tau_c = \frac{\lambda}{V_s}$ is the time between collisions with the beads in the porous layer, and $\lambda$ is the swimmer mean free path. The timescale $\tau = \tau_c + \tau_R$ is the total porous travel time, including the residence time $\tau_R$ that a swimmer spends at an obstacle. These parameters were recently measured experimentally for *C. reinhardtii* [36] (see Table 1). To express the system (1) in terms of concentrations only, we note that the mean number of cells in regions $i = u, p, l$ can be written as $\bar{N}_i = Ah_i\bar{c}_i$, where, as above, A is the cross-sectional area of the cuvette, and $h_i$, $\bar{c}_i$ are the height and mean concentration in region $i$, respectively. Substituting into (1), we thus obtain, dividing both sides by the respective $h_i$,

$$\frac{d\bar{c}_l}{dt} = -\alpha\,\bar{c}_l\,, \tag{3a}$$

$$\frac{d\bar{c}_p}{dt} = \beta\,\bar{c}_l - \gamma\,\bar{c}_p\,, \tag{3b}$$

$$\frac{d\bar{c}_u}{dt} = \delta\,\bar{c}_p\,, \tag{3c}$$

where we have defined the upswimming rate constants $\alpha = V_s/h_l$, $\beta = V_s/h_p$, $\gamma = V_{eff}/h_p$ and $\delta = V_{eff}/h_u$. Equation (3a) has immediate solution $\bar{c}_l = k_0 e^{-\alpha\,t}$, where $k_0$ is a constant. The system of equations (3) can then be solved analytically by substituting this solution into (3b), and the resulting solution (e.g. by using the integrating factor $e^{\beta\,t}$) into (3c). Applying the initial conditions $\bar{c}_l(0) = \bar{c}_l{}^0$, $\bar{c}_p(0) = \bar{c}_p{}^0$ and $\bar{c}_u(0) = \bar{c}_u{}^0$, where $\bar{c}_i{}^0$ represent the initial average concentrations in the three regions, we find:

$$\bar{c}_l(t) = \bar{c}_l{}^0 e^{-\alpha\,t}\,, \tag{4a}$$

$$\bar{c}_p(t) = \bar{c}_l{}^0 \frac{\beta}{\gamma-\alpha} e^{-\alpha\,t} + \left(\bar{c}_p{}^0 - \bar{c}_l{}^0 \frac{\beta}{\gamma-\alpha}\right) e^{-\gamma\,t}\,, \tag{4b}$$

$$\bar{c}_u(t) = \bar{c}_u{}^\infty - \bar{c}_l{}^0 \frac{\beta\delta}{\alpha(\gamma-\alpha)} e^{-\alpha\,t} - \left(\bar{c}_p{}^0 - \bar{c}_l{}^0 \frac{\beta}{\gamma-\alpha}\right)\frac{\delta}{\gamma} e^{-\gamma\,t}\,, \tag{4c}$$

where we have defined the long-time concentration in the upper region as

$$\bar{c}_u{}^\infty = \bar{c}_u{}^0 + \bar{c}_p{}^0 \frac{\delta}{\gamma} + \bar{c}_l{}^0 \frac{\beta\delta}{\alpha\,\gamma} = \bar{c}_u{}^0 + \bar{c}_p{}^0 \frac{h_p}{h_u} + \bar{c}_l{}^0 \frac{h_l}{h_u}\,, \tag{5}$$

and where, recalling the definitions of the constants $\alpha$, $\beta$, $\gamma$ and $\delta$, we have re-written $\bar{c}_u{}^\infty$ in terms of the heights of the regions. Thus, it is clear from equation (5) that, in this simple model, the long-time (maximum) concentration in the upper region occurs when all swimmers from the porous and lower regions have concentrated themselves into the upper region.

We also consider the 'mesh-only' case (without a porous layer of beads). The derivation, shown in Appendix B, is similar and provides the temporal evolution of the mean concentrations as

$$\bar{c}_l^m(t) = \bar{c}_l^{m0} e^{-\alpha\,t}\,, \tag{6a}$$

$$\bar{c}_u^m(t) = \bar{c}_u^{m\infty} - \bar{c}_l^{m0} \frac{\eta}{\alpha} e^{-\alpha\,t}\,, \tag{6b}$$

where the superscript '*m*' denotes concentrations in the mesh-only case, and we have defined the rate constants $\alpha = V_s/h_l$, which is as in the porous layer model (but takes a slightly different value because of the different value of $h_l$, see <span style="color:red">Table 2</span> below), and $\eta = V_s/h_u$. For the mesh-only case the concentration in the upper region at long times is given by

$$\bar{c}_u^{m\infty} = \bar{c}_u^{m0} + \bar{c}_l^{m0}\frac{\eta}{\alpha} = \bar{c}_u^{m0} + \bar{c}_l^{m0}\frac{h_l}{h_u}. \tag{7}$$

This corresponds to the concentration in the upper region occurring when all microalgae have swum into it from the lower region.

In the upper region we observe that cells accumulate strongly at the surface. To describe this, we can use a simplification of the Williams and Bees model [9]. By virtue of assumptions v)-viii) above, as shown in Appendix A, the full swimmer conservation equation in the Williams and Bees model simplifies to:

$$\frac{\partial c_u}{\partial t} = -\nabla \cdot [\, V_s\, c_u\, \boldsymbol{k} - D\, \nabla c_u \,], \tag{8}$$

where $\boldsymbol{k}$ is a unit vector pointing upwards and $D$ is the diffusivity, approximated as isotropic, by assumption vii) (see Appendix A for more details). By assumption i), we have a steady state, so that (6) implies

$$V_s\, c_u\, \boldsymbol{k} - D\, \nabla c_u = const., \tag{9}$$

where $\boldsymbol{k}$ is a unit vector pointing upwards. Imposing a no flux condition at the upper boundary (flat for simplicity, assumption viii) requires $(V_s\, c_u\, \boldsymbol{k} - D\, \nabla c_u) \cdot \boldsymbol{k} = 0$ on $z = h$, so that equation (7) becomes

$$\frac{\mathrm{d}c_u}{dz} = \frac{V_s}{D} c_u \,, \tag{10}$$

which integrates to

$$c_u = k_1\, e^{\frac{z}{l_p}}, \tag{11}$$

where we have defined a characteristic phototactic accumulation lengthscale $l_p = \frac{D}{V_s}$, and where $k_1$ is a constant. To find the latter, we use the fact that the average background concentration is given by $\bar{c}_u$, that is, taking $z = 0$ at the bottom of the upper region and $z = h_u$ at its top, $\bar{c}_u = \frac{1}{h_u}\int_0^{h_u} c_u\, dz$. Thus, integrating equation (9) gives $k_1 = \bar{c}_u\, (e^{h_u/l_p} - 1)^{-1}\, h_u/l_p$, so that finally the distribution in the upper region is given by

$$c_u(z, t) = \bar{c}_u(t) \frac{h_u}{l_p} \frac{e^{\frac{z}{l_p}}}{e^{\frac{h_u}{l_p}} - 1} \quad , \tag{12}$$

where that the mean concentration as a function of time, $\bar{c}_u(t)$, is provided by equation (4c).

**Table 1. Essential model parameters for the mesh+beads case.** Values were obtained from direct measurements of our experimental system or literature values for the swimming parameter of *C. reinhardtii* grown under identical conditions.

| Parameter | Symbol | Units | Value | Reference |
|---|---|---|---|---|
| Mean swimming speed of *C. reinhardtii* | $V_s$ | $cm/s$ | $80 \times 10^{-4}$ | [34] |
| Rotational diffusivity of *C. reinhardtii* | $D_R$ | $s^{-1}$ | 0.4 | [34] |
| Effective diffusivity of *C. reinhardtii* | $D = \frac{V_s^2}{D_R}$ | $cm^2/s$ | $1.6 \times 10^4$ | [34] |
| Mean free path in porous layer | $\lambda$ | $cm$ | $125 \times 10^{-4}$ | This work |
| Collision time in porous layer | $\tau_c = \frac{\lambda}{V_s}$ | $s$ | 1.56 | This work |
| Residence time at obstacle in porous layer | $\tau_R$ | $s$ | 1 | [36] |
| Mean run time | $\tau = \tau_c + \tau_R$ | $s$ | 2.56 | [36] |
| Mean distance on obstacles | $l_{obs}$ | $cm$ | $30 \times 10^{-4}$ | [36] |
| Lower reservoir region height | $h_l$ | $cm$ | 0.212 | This work |
| Porous region height | $h_p$ | $cm$ | 0.378 | This work |

| Upper harvest region height | $h_u$ | $cm$ | 0.422 | This work |
|---|---|---|---|---|
| Initial mean concentration of suspension in the lower region | $\bar{c_l}^0$ | $cells\ cm^{-3}$ | $1.20 \times 10^6$ | This work |
| Initial mean concentration of suspension in the porous region | $\bar{c_p}^0$ | $cells\ cm^{-3}$ | $1.18 \times 10^6$ $(\bar{c_p}^0 = \bar{c_u}^0)$ | This work |
| Initial mean concentration of suspension in the upper region | $\bar{c_u}^0$ | $cells\ cm^{-3}$ | $1.18 \times 10^6$ | This work |
| Phototactic lengthscale | $l_p = \dfrac{D}{V_s}$ | $cm$ | 0.02 | This work |
| Upswimming rate 1 | $\alpha = \dfrac{V_s}{h_l}$ | $s$ | $3.8 \times 10^{-3}$ | This work |
| Upswimming rate 2 | $\beta = \dfrac{V_s}{h_p}$ | $s$ | $2.12 \times 10^{-2}$ | This work |
| Upswimming rate 3 | $\gamma = \dfrac{V_{eff}}{h_p}$ | $s$ | $1.60 \times 10^{-2}$ | This work |
| Upswimming rate 4 | $\delta = \dfrac{V_{eff}}{h_u}$ | $s$ | $1.43 \times 10^{-2}$ | This work |

## 3    Results

### 3.1    Initial condition for the lower region: a bottom-standing plume

Prior to considering the effect of light on a suspension of *C. reinhardtii* placed in the cuvette, we will consider the initial condition of the suspension in the lower reservoir region, which will be the same starting point for all subsequent experiments. With the blue LED light off, microalgae were mixed into the cuvette and the suspension was allowed to stabilize in the presence of only red illumination from the side (see Figure 1a), which does not elicit a phototactic response (SM Video 1) [37]. The suspension images and profiles are shown as a time-series in Figure 2: over a few minutes, the suspension (initial concentration ∼ 1 million/mL) settles into a distribution where the majority of cells reside at the bottom of the cuvette; a steady distribution is observable beyond 4 minutes. The gradient in concentration already visible for the concentration profile at t = 0 is due to a lag in transferring the cuvette to the imaging setup after mixing: some settling has already occurred at the first instance of imaging. The steady distribution observed beyond 4 minutes coincides with the "bottom-standing plumes" first described by Kessler [38]. These come about because concentration fluctuations arising in the initially well-mixed suspension collectively sink because individual cells are negatively buoyant (*C. reinhardtii* is 10% denser than the surrounding medium [34]). This sinking drives a local shear flow, which, because of gyrotaxis, reorients cells to swim towards the line where the siniking fluctuation occurred, generating thread-like structures, known as plumes. Because the plumes sink faster than the cells can swim upwards, they deliver cells to the bottom of the container. Since there is no net flow imposed in the cuvette, to the downflow caused by sinking plumes corresponds an upflow. The latter will advect swimming cells upward, but these will also, once more, be gyrotactically reoriented by the downwelling plume to swim into it, and be advected downwars. And so this process repeats, resulting in a stable recirculating structure: the bottom-standing plume. We note that for *C. reinhardtii*, the bottom-standing plume structures are not prominently visible in the images of the cuvette (plume dynamics can be seen in SM Video 1). This is likely because, for this species, the bottom-standing plume wavelength is if the same order as the cuvette width, so that plumes arise at the cuvette walls. At similar concentrations the plumes are clearly visible for other species, such as *Chlamydomonas augustae* [2] (see also Figure 1a in [7]).

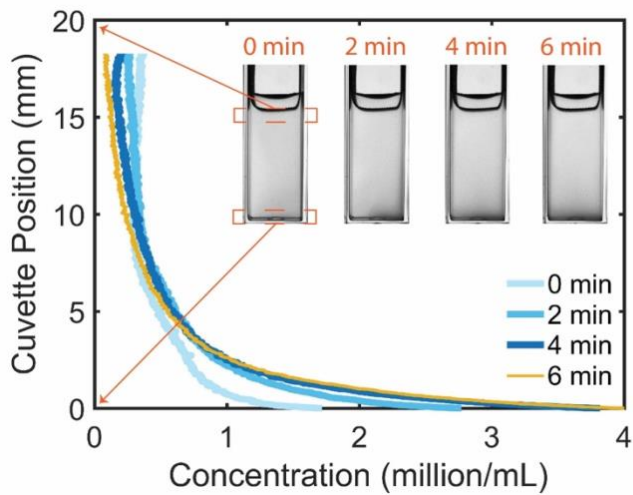

**Figure 2.** In the absence of phototactic illumination from above microalgae redistribute over the height of the cuvette. A bottom-standing plume (as broad as the cuvette is wide) can be seen to stabilize after 4 min.

## 3.2   Free surface: bulk photogyrotactic instabilities

We consider here the effect of light on a suspension of microalgae in a cuvette when the surface of the suspension is free (the metal mesh applied in the next section has been raised above the surface). This experimental scenario can be seen in SM Video 2, stills of which are shown as the sequence in Figure 3. Initially the blue LED illumination is switched off and the suspension is distributed in a bottom-standing plume, as described in the previous section. Then the LED is switched on, and the cells in suspension phototactically respond to the light, migrating upwards toward the surface (Figure 3, t = 3 min). Concomitantly, instabilities arise throughout the suspension, resulting in meandering plumes (Figure 3, t = 3 − 5 min). These are of photogyrotactic origin, as discussed below. In the span of ∼ 6 minutes phototactic migration appears to have delivered many swimmers to the surface, leaving the bulk of the suspension depleted. This surface accumulation is gravitationally unstable because of the negative buoyancy of surface-accumulated cells: it results in the formation of a plume instability seen to originate from the middle of the meniscus of the suspension surface (Figure 3, t = 9 min). The plume structure wiggles around but, once formed, is dynamically stable (Figure 3, t = 18 min), delivering cells to the bottom of the container. Once they reach this, the microalgae migrate back up to the surface to join the plume, and so forth. When the light is switched off (Figure 3, t = LED Off + 17 sec, + 3, 6 min), the phototactic migration toward the surface stops and the surface accumulation sinks as a broader, non-meandering plume. This takes the cells to the bottom of the cuvette, where they once more settle as a bottom-standing plume.

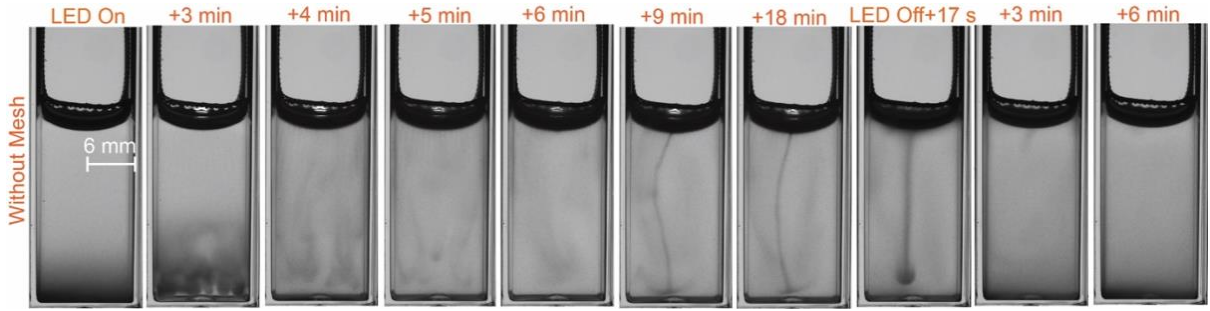

**Figure 3.** Free surface: photogyrotactic dynamics of a suspension of *C. reinhardtii* microalgae in a cuvette illuminated from the top by a blue LED, and dynamics when the LED is switched off (last three stills).

### 3.3    Mesh: new phototactic structures

In this section we consider the case of a metal mesh immersed at the top of the microalgal suspension. As described in the methods, the pore size of the mesh is 350 µm, so individual microalgae (~ 5 µm in diameter) easily swim through it. A typical experiment is shown in SM Video 3, stills of which are presented as a sequence in Figure 4a (top row). As for the free surface case, blue LED illumination is initially off, and the suspension is distributed in a bottom-standing plume (Figure 4a, first still). The LED is then switched on and microalgae migrate upward in response to the light (Figure 4a, t = 3 min). The response is broadly similar to the free surface case, but there are some interesting differences. One such striking difference is that the mesh creates a pattern of light and shadow to which the microalgae visibly respond photogyrotactically, forming accumulations ('phototactic curtains'), see Figure 4c. The average width of phototactic curtain feature is $630 \pm 53$ µm, nearly twice the mesh pore size, showing that the curtains are not the result of shadowing by the mesh, but genuine phototactic structures originating from the response of the microalgae to the local light profile. As in the case of a free surface, when the density of cells phototactically accumulated at the surface becomes too high, a plume of dense cells forms and sinks. However, viscous resistance caused by the mesh pores prevents the plume from completely sinking beyond the mesh, and instead a cloud-like plume structure is seen to be trapped, hovering above the mesh (Figure 4a, t = 9, 18 min). Not all the plume-cloud is trapped, negative buoyancy is sufficient to cause some of it to escape through the mesh forming a meandering secondary plume, similar in appearance to the one observed in the free surface case (Figure 4a, t = 9, 18 min). While the light is on, these structures appear dynamically stable. As the light is switched off, however, the curtains and cloud structure disperse, cells sink through the mesh, and the escaped plume sinks down straight, again similarly to free surface case (Figure 4a, t = LED Off + 19 sec, + 3, 6 min). This emphasizes the stabilizing influence of phototaxis: none of the observed structures could be possible in the absence of the light. Both mesh and light are critical for supporting the plume-cloud.

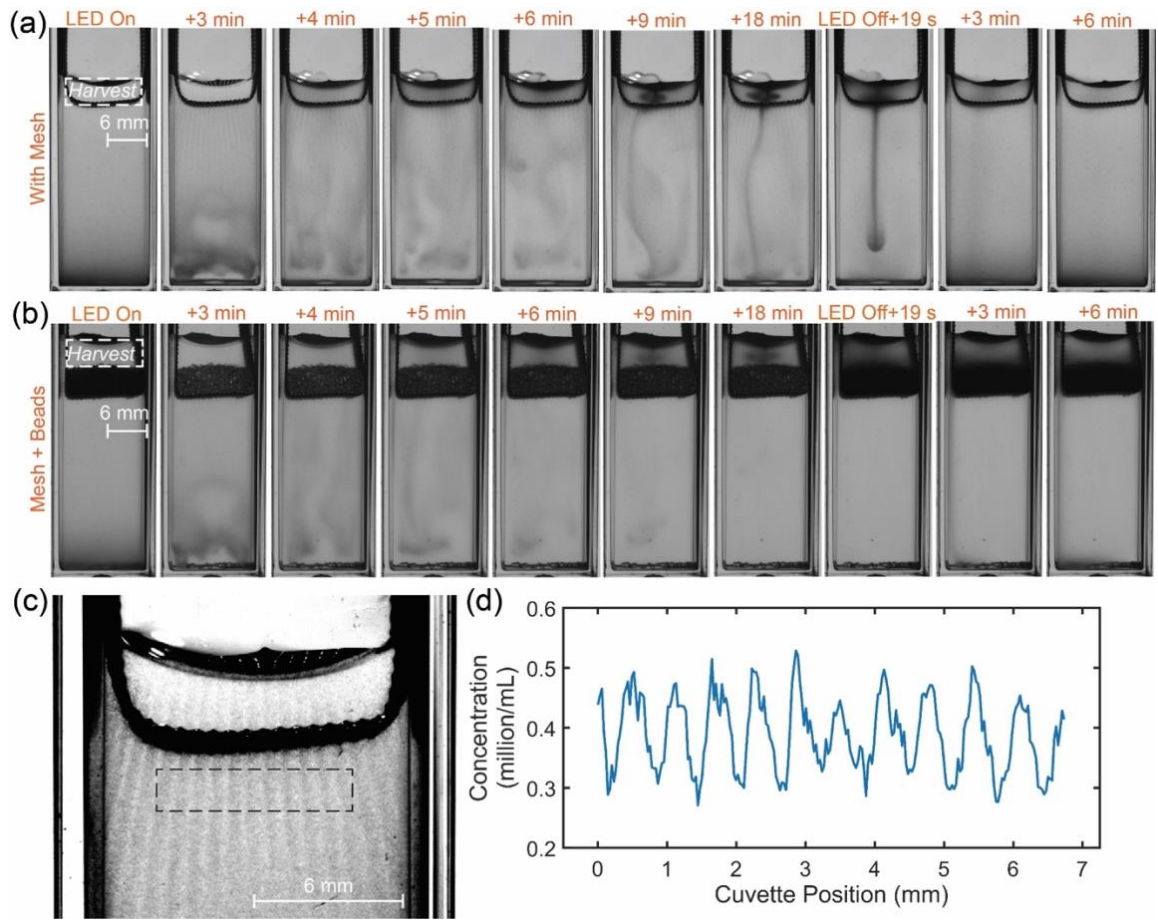

**Figure 4.** Microalgal suspension dynamics for the case of: **(a)** a mesh; **(b)** mesh+beads (400 mg) placed at the top of the suspension. The dynamics is similar, but there are important differences. Significantly a cloud-like plume is completely trapped between the porous layer and top surface, while in the case of the mesh, it can leak as a thin plume to the suspension below. **(c)** Phototactic 'curtain' pattern formed by the accumulation of cells in response to the light and shadow pattern generated by illumination falling on the mesh. **(d)** Curve showing the concentration of algae across the phototactic curtains (mean feature width = $630 \pm 53$ μm).

## 3.4  Porous layers: stabilization of phototactic structures and concentration gain

We next turn to the case where a porous layer is placed on top of the suspension. As described in the Methods, the porous layer consists of glass beads overlaid onto a metal mesh (the same as was used in the previous section). The beads are around $425 - 600$ μm in diameter, which results in interparticle spacings $\sim 50 - 200$ μm (from microscopic observation). Thus, individual algae $\sim 5$ μm in diameter can swim through the porous layer. We studied the effect of light on suspensions of microalgae overlaid with porous layers, quantified by the weight of the beads placed on the mesh. A typical experiment with a layer weighing 400 mg is shown in <span style="color:red">SM Video 4</span>, and stills from this video are presented in <span style="color:red">Figure 4b</span>. As in previous cases, the LED light is initially off and the suspension is distributed as a bottom-standing plume (<span style="color:red">Figure 4b</span>, first still). When the LED is switched on, the initial suspension dynamics are similar to the mesh-only case (<span style="color:red">Figure 4b</span>, $t = 3 - 6$ min), displaying instabilities as the microalgae respond to the light (but with no curtains visible). However, for this case, we were also able to observe clusters of cells swimming upwards as waves in response to the light, see <span style="color:red">Figure 5a</span> for an example. Averaging over five such waves, we found them to have a mean speed of $190 \pm 60$ μm/s. This is faster than mean swimming speed of individual algal cells, 80 μm/s [34], possibly as a result of advection by upwelling fluid in the lower region of the cuvette generated by the photogyrotactic suspension dynamics. The large deviation in the speed of the waves could also be due to the interaction of the waves with other photogyrotactic structures and up/downwelling flows in the suspension. Once cells have had time to accumulate in the harvest region and on the surface of the suspension, a plume-cloud structure originating at the low point of the meniscus forms above the porous layer (Figure 4b, t = 9, 18 min). The plume-cloud appears more diffuse than in the mesh case. The time taken for the plume-cloud to arise in 10 out of 12 experiments used for the analysis is between 6 − 10 min from when the LED light is switched on, as shown in SM Figure 3. Unlike the case of the mesh, the plume does not leak through the porous layer into the suspension: in the presence of light, the viscous resistance offered by the porous layer is sufficient to stabilize the plume-cloud. Instead of sinking the plume-cloud is observed to gradually expand into the upper region. Figure 5b charts this expansion. The lateral extent of the plume structure increases the most between 7-9 min after the LED has been switched on, when the plume begins to drop and propagate along the porous layer. After that the plume-cloud achieves a steady structure, probably as a result of balance between influx of cells from the surface, where the plume originated at the low point of the meniscus, and loss to the edges of the harvest region (and resorption to the suspension surface by upswimming).

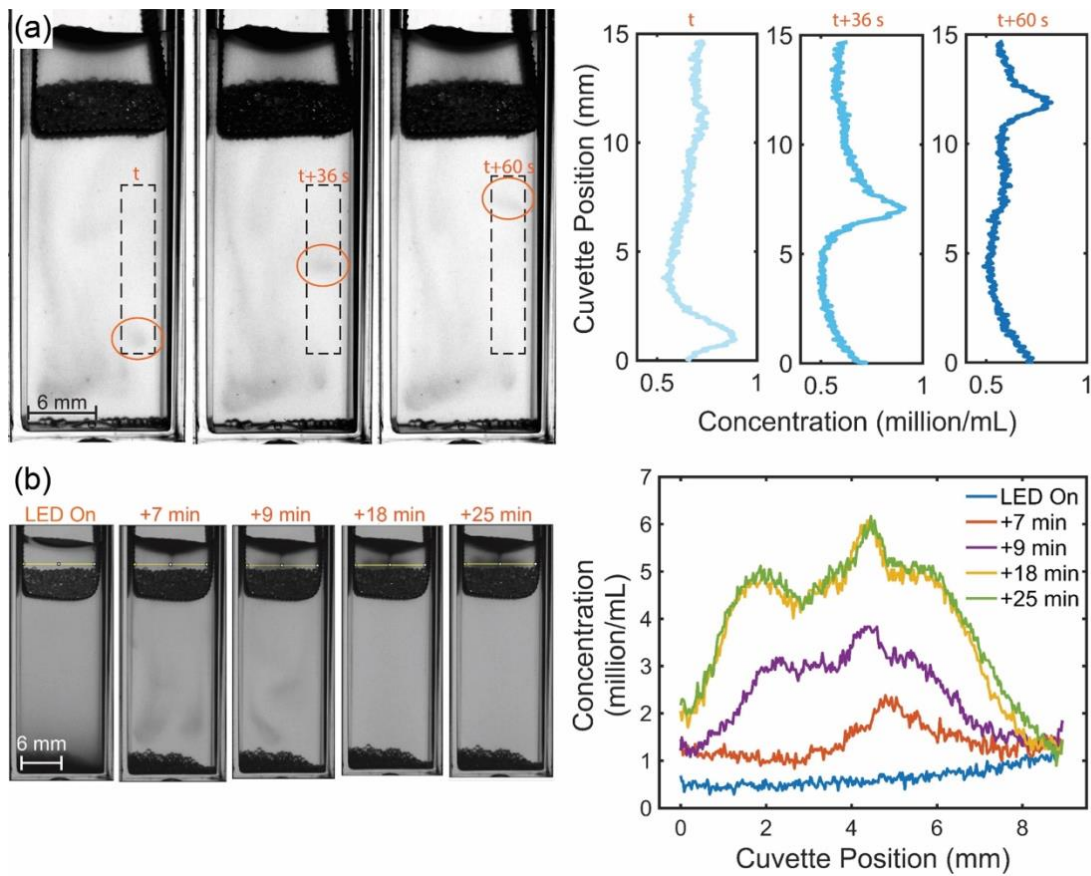

**Figure 5.** Photogyrotactic dynamics of swimming algae in the mesh+beads case. (a) Stills of the algal clusters in the lower region moving upwards as waves with speed ~180 μm/s. (b) Lateral expansion of the trapped algal plume-cloud formed in the upper harvest region. Its density is seen to increases over time as it expands.

In view of quantifying microalgal concentration in the upper harvest region above the mesh or the mesh+beads porous layer, it is instructive to chart the evolution of the average concentration of the suspension in this region (SM Figure 2). To identify a porous layer thickness that would not leak into the suspension below, we considered layers of several weights in trial experiments presented in SM Figure 4. We found a general qualitative trend that was similar for all cases: the concentration grows as the light is switched on, and then saturates to a constant value. We focus here quantitatively on the mesh-only case and the 'minimally-leaky' mesh+beads (400 mg) case, shown for three repeats in Figure 6a and 6b, respectively. The averaged profiles are shown in Figure 6c. This makes it clear that the concentration in the upper harvest region of both the mesh and mesh+beads cases, following a dip in concentration due to phototactic accumulation of cells to the upper surface, grows after the LED is switched on and then tends to saturate. The mesh case, however, saturates earlier, probably because of losses to the lower region, such as the plume visible in Figure 4a (9 min). Another interesting quantitative difference between the two cases is the initial rate of concentration, which appears slightly larger for the mesh case. This indicates the concentration process is initially slower when a porous layer is present, than in its absence. As discussed below using the essential model, this makes sense in terms of the microalgae having to make their way through the porous layer, which reduces the swimming speed that sets the concentration rate. The difference in swimming speed will also affect the average time it takes to form the plume, which was measured to be $7.3 \pm 0.6$ min for the mesh case, while it is $8.7 \pm 1.5$ min for mesh+beads (SM Figure 3). When the LED is switched off (Figure 6a-c inset), the concentration in the upper harvest region is seen to rise briefly before steadily falling. This is because, with the light off, the concentrated algal suspension in the harvest region no longer responds phototactically and cells accumulated to the surface are released, sinking down as dense fluid. The increase in concentration due to the cells coming off the surface shows that our measurements likely underestimate the concentration in the upper harvest region because of cells 'hidden' at the surface. This could account, at least in part, for discrepancies with model predictions discussed below. For our setup, the time after switching the LED off is optimal for harvesting the suspension, yielding a harvest concentration $\approx 5$ million/mL (gain $\approx 4.2$ compared to the initial concentration) for the mesh+beads case, as compared to $\approx 4$ million/mL cells for mesh-only (gain $\approx 3.6$). This highlights the advantage of concentrating using a porous layer. The latter also slows down the rate at which the cells sink back through to the lower region, which depends on the layer thickness.

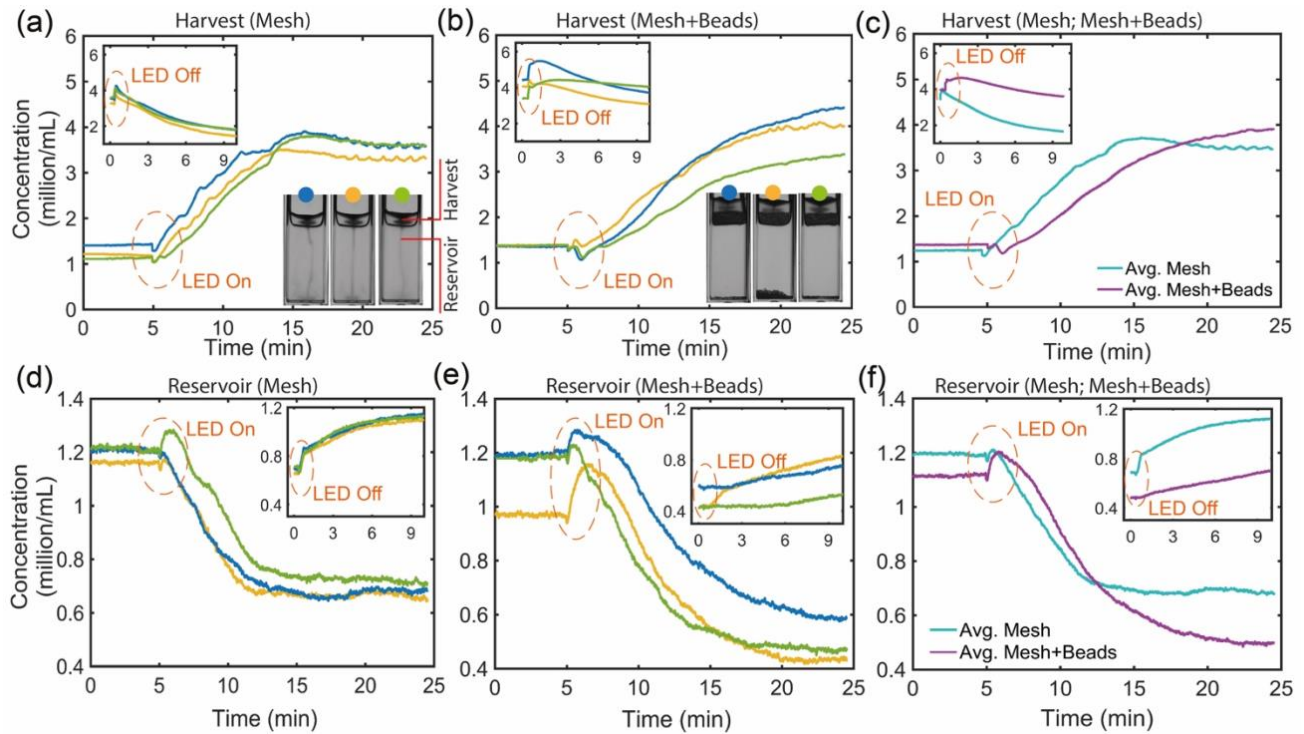

**Figure 6.** Temporal concentration profiles in the upper harvest region above a mesh or mesh+beads, after the LED is switched on and off (insets), as indicated. (a) Three repeats for microalgae phototactically concentrating above the bare

mesh. Inset: concentration rise and decay after the LED is switched off. (b) As in (a), but for the mesh+beads (400 mg) case. (c) Time point average of the concentration profiles shown in (a) and (b). The concentration for the mesh-only case saturates ~ 10 min after the LED is switched on, whereas it keeps on increasing towards a higher saturation concentration in the mesh+beads case. When the LED is switched off, the concentration initially increases and then decays, for the reasons discussed in the text. (d) Three repeats for profiles in the lower region for the mesh only case. Inset: concentration rise after the LED is switched off. (e) As in (d) but for the mesh+beads case. (f) Time point average of the concentration profiles shown in (d) and (e). For the mesh-only case the profile decays to a constant value sooner than mesh+beads. The inset shows how the concentrations for both cases rise in the lower region after the LED is switched off, with a greater rise for the more leaky mesh-only case.

Also shown in Figure 6 are profiles charting the temporal evolution of concentration in the lower reservoir region. As for the upper region, we have measured triplicate repeat profiles for the mesh (Figure 6d) and mesh+beads (Figure 6e), and also evaluated averaged profiles (Figure 6f). We see that, after the LED is switched on, the concentration for the mesh and mesh+beads falls, as phototactic swimming into the upper regions depletes the lower region of cells. However, the depletion appears to saturate, and to a higher concentration in the case of mesh-only, reflecting the greater leakiness of the mesh, as discussed below. Insets in Figure 6d-f display how, with the LED off, the concentration in the lower region rises due the influx of cells sinking from the upper regions.

### 3.5 Essential model predictions

We have developed a simple model to capture the essential features of the phototactic concentration dynamics, and evaluate it here using parameters for *C. reinhartii* concentrated using a mesh+beads setup, as shown in Table 1. In Figure 7a, the model prediction using equation (4c) for the average concentration of cells $\bar{c}_u(t)$ in the upper harvest region is shown as a function of time (the concentration process starts at time t = 0, 'LED on'). Qualitatively, the predicted behaviour is as in the experimental curves (Figure 6c), with the concentration initially rising and then saturating. However, quantitatively, the concentration values predicted by the essential model are much larger than those seen experimentally. Indeed, using equation (5) and the parameters in Table 1, the essential model predicts saturation to a long-time concentration $\bar{c}_u^\infty = 8.2 \times 10^6$ cells/ml. This is of the same order of magnitude as, but approximately double what we observe experimentally ($\approx 4 \times 10^6$ cells/ml). Part of the discrepancy is because, as mentioned above, experimental concentration curves underestimate the concentration in the upper reservoir because of swimmers phototactically accumulated and 'hidden' at the surface. Another possible reason is that the essential model unrealistically ignores mechanisms causing losses: as illustrated by equation (5), $\bar{c}_u^\infty$ corresponds to the concentration obtained when all the swimmers from the porous and lower regions swim to the upper region and do not leave it thereafter. In reality, swimmer diffusion will cause cells to be transferred from the upper to the porous region, particularly at longer times when concentration gradients between the regions are large. Another possibility not accounted for by the essential model is that, if swimmers respond to gradients of light (as opposed to just its direction, as assumed in the model), the denser suspension of swimmers in the upper region shades the region below, changing the light gradient and reducing the phototactic speed of swimmers below, and thus the rate of accumulation. Figure 7a also shows the model prediction for the concentration in the upper region for the mesh-only case. As in experiment, this is seen to initially rise steeper and saturate at a lower value than the case of mesh+beads; numerically, however, the predicted concentrations ($\bar{c}_u^\infty = 7.5 \times 10^6$ cells/ml) are approximately double what we measured experimentally. This is for the same reasons as for the mesh+beads case, and additionally in the mesh-only case there are also losses due to the plume leaking through the mesh, as we have shown (Figure 4a, 9 min). The faster rise in concentration observed for the mesh-only case compared to mesh+beads, is due to the difference in upswimming rates in the two cases: for mesh+beads, swimmers are slowed down when they swim through the porous layer. The essential model also allows the prediction of the concentrations in the porous and lower regions, respectively $\bar{c}_p(t)$ and $\bar{c}_l(t)$, which are shown in **Figure 7a**. The lower region concentration is seen to decay exponentially to zero, as swimmers evacuate the lower region by phototactic upswimming. The lower region decay predictions overlap for the mesh and mesh+beads cases, so they are not separately visible in the figure (the prediction equation is the same for these cases, and parameters are practically identical). We can compare these predictions with decay with the experimentally determined concentration profiles. As observed in the previous section, these also decay with time, but not to zero: they saturate to a fixed value (**Figure 6f**), with the mesh-only case reaching a lower value than mesh+beads due to the greater leakiness of the mesh. The essential model fails to predict this saturation and the important difference between the two cases, demonstrating the need to model diffusive transfer and leaking plumes between the regions, and/or a reduction of the phototactic speed. For the porous region, the essential model predicts that the concentration, $\bar{c}_p(t)$, initially rises, due to influx from the lower region outpacing losses to the upper region, and eventually decays to zero. It was not possible to optically image the microalgae in the porous region and obtain the concentration there, so we cannot make a comparison with the essential model prediction in this case.

Assuming phototaxis and diffusion processes are dominant in the upper region, and that these occur faster than the accumulation from the porous region, we can also use the spatial extension of the essential model to chart the distribution of swimming algae in the upper region, which is provided by equation (11). We note that, since this model does not fully account for photogyrotaxis, the predictions are only strictly valid prior to the formation of the plume off the upper surface, which we know from experiment occurs $\approx 9$ minutes after turning the light on. In **Figure 7b** the distribution of swimmers is charted at different points in time (prior to plume formation), predicting that the suspension becomes increasingly top-heavy as time progresses. This accumulation, with concentrations reaching $\sim 10^8$ cells/ml close to the upper boundary, is unstable against its own negative buoyancy, and eventually results in the formation of the plume we observe experimentally. As it is not possible to accurately image the accumulation of cells around the meniscus, we did not experimentally quantify the spatial concentration distribution in the upper region. However, the increasing accumulation of swimmers at the surface is clearly discernible in our image sequences, see **SM Video 4**. The model predicts that the cells accumulate strongly at the top of the upper region, with no sizeable concentration below a certain height. Instead, our image sequences reveal that there is also a nonzero concentration in the bottom of the upper region (indeed that is what we have measured to obtain **Figure 6a-c**). This could be accounted for by losses from the accumulation at the surface to the edge of the cuvette, which are not considered in our model.

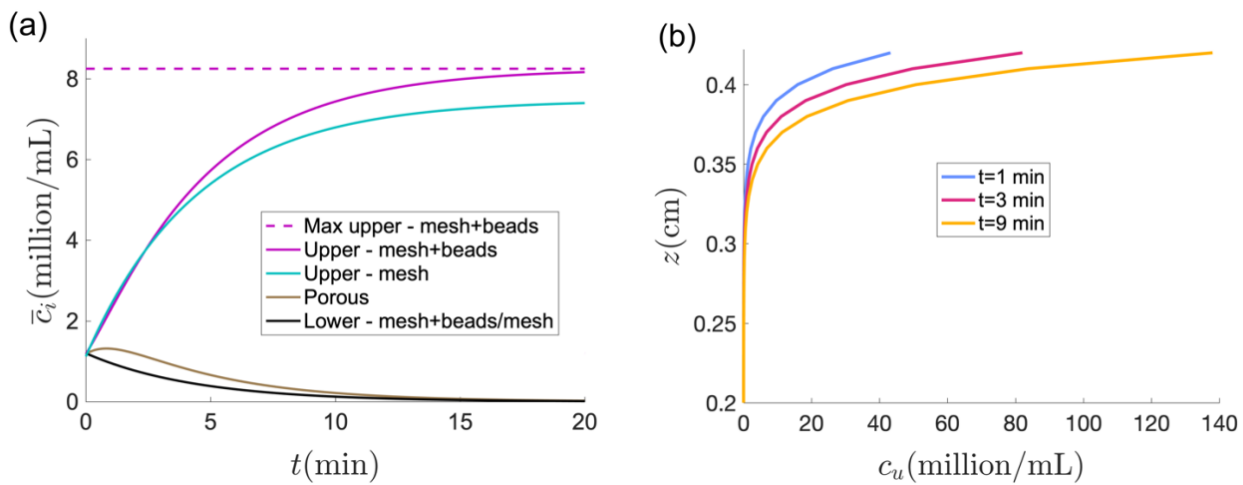

**Figure 7.** Predictions of the essential model. (a) Average concentrations as a function of time since 'LED on' at t = 0 in the upper, porous, and lower regions for the mesh+beads case, and upper and lower regions for the mesh-only case. For both cases, phototactic concentration causes the lower region to evacuate and the upper region to fill up with swimmers, up to a maximum limit, as discussed in the text. For clarity, this limit, shown as a dotted line for the mesh+beads case, is not shown for the mesh-only case. (b) Spatial swimmer concentration profiles in the upper region at different times, as shown. The suspension becomes increasingly top-heavy. The plot starts at z = 0.2 cm to make the profiles more evident (the concentration predicted below this level is 0 cells/ml).

## 4    Discussion

We have shown how light from above can trigger instabilities and upwards migration in an initially quiescent suspension of *C. reinhardtii* microalgae within a rectangular cuvette. By imaging, we qualitatively and quantitatively studied for the first time this migration in the following cases: when a permeable metal mesh is placed at the top of the suspension; when porous layers of beads are overlaid onto the mesh; in the absence of any mesh or layer on the surface. In the latter case, light was seen to drive photogyroactic instabilities in the bulk of the suspension and upwards migration of the cells to the surface, from which, eventually, a plume structure was seen to arise. A similar phenomenology was observed when a mesh was present, except in this case the plume from the surface was partially trapped by the mesh, later giving rise to a secondary plume. By trapping the plume, the mesh allows the concentration of cells in the upper region of the cuvette (also termed 'harvest region'), but this is a leaky process. However, when a porous layer of glass beads is overlaid onto the mesh, it is possible to stably concentrate the suspension in the upper harvest region while the light is switched on: the plume from the surface is trapped with minimal leakage. We have charted how the mean concentration in the harvest region varies with time for the case of a mesh with a layer of beads of different weights (thicknesses), showing that a ≈ 4-fold concentration is possible for the thickest layer weighing 400 mg ([Figure 6c]). Critically, we have demonstrated that it is the unique combination of light and a moderately thick porous layer of beads that makes the photogyrotactic concentration of cells possible. Without the beads the accumulation of microalgae in the harvest region is leaky. When the light is switched off, all photogyrotactic structures fall apart, and the microalgal population sinks back down to the initial quiescent state.

We can discuss our findings in terms of what is known about the phototactic and photogyrotactic behaviour of microalgal suspensions. When the LED is switched on, the suspension responds visibly in seconds, similarly to what has been measured for populations responding to light from an optical fibre [24], and corresponding to the time scale for *C. reinhardtii* to perceive light and turn to swim towards it by controlling their flagellar beat [25]. Subsequent to this initial response, the suspension displays instabilities in cell concentration and flow. Some groups of cells rise, moving as waves drifting at the swimming speed of the algae or above, probably advected by upwelling flow; others form stretching plumes; others still sink. This complex behaviour is the result of the interplay of the phototaxis and gyrotaxis of the population, coupled with the fluid dynamics of a negatively buoyant suspension. In the absence of a full photogyrotactic model, whose development is beyond the scope of this paper, it is not possible to account for these observed patterns quantitatively. A lower bound estimate of the timescale for accumulation to the surface leading to the formation of a plume there can, however, be obtained by considering the time for cells to swim straight up to the surface at the maximum phototactic speed. For the mesh+beads (400 mg) case, the mean swimming speed of the microalgae in the lower and upper regions, with heights $h_l$ and $h_u$, respectively, is $V_s$, while it is $V_{eff}$, as given by equation (2), in the porous region with height $h_p$. The time to reach the surface is then t ∼ $(h_l + h_u)/V_s + h_p/V_{eff} ≈ 6$ min, using parameters in Table 1. For the mesh-only case, there is no porous layer so that $t_m ∼ (h_l + h_u)/V_s ≈ 5$ min, using parameters in Table 2. These values are not too far from the ∼ 9 (7) min it takes for a plume instability to develop from the surface in the mesh+beads (mesh) cases (SM Figure 3). This suggests, as is clear from our concentration data in the harvest region (Figure 6), that, in spite of the instabilities observed, phototaxis drives a net flux of cells upwards through the porous layer towards the harvest region, where cells accumulate at the surface. Here they distribute, with maximum concentration at the surface. The concentration becomes increasingly top-heavy (as predicted by our essential model, see Figure 7b) and eventually a plume instability develops driven by the negative buoyancy of the suspension. The plume then drops towards the porous layer, but its negative buoyancy is not great enough to sink through it. Instead, the plume is trapped in the harvest region forming a toroidal 'plume-cloud' fed from the surface, whose size expands with time to the edge of the harvest region (see Figure 5b). Here it appears to stabilize, possibly due to a balance between cell gain from the surface and loss to the suspension at the edges of the cuvette. The surface accumulation and plume-cloud, and all the structures in the lower region, collapse within a few seconds of switching the LED light off. In particular, the concentrated suspension in the harvest region sinks right through the porous layer, though this takes some time for the thick (400 mg) mesh+bead layer. This collapse of the suspension structures demonstrates the essential role of phototaxis in dramatically altering the stability thresholds of the active suspension: none of the structures we have observed can exist without light.

Our essential model provides a qualitative picture of how the average concentration changes in the upper, porous and lower regions, and gives concentration values which agree in order of magnitude with what we have measured. Comparison with experiment, however, reveals that the model fails to quantitatively describe the saturation of the upper and lower concentrations. This in part because our measurements in the upper region underestimate the concentration (missing cells accumulated at the surface). However, as evidenced by the failure of the model to predict saturation in the lower region (compare Figure 7b and Figure 6d), it is likely that quantitative agreement is not possible because critical processes have not been modelled, such as diffusive exchanges between reservoirs and/or shading effects of the cell concentration in the upper region on the phototactic speed. For the upper region, the model was applied to predict a top-heavy distribution of cells, as is observed in our image sequences. The model, however, does not reproduce the concentration of cells visible in the bottom part of the upper region, probably due to a neglect of losses from the surface accumulation at the edge of the cuvette. The model is further limited to the description of the phototactic concentration prior to the formation of the plume-cloud, whose quantitative dynamics require a fully photogyrotactic description. Future studies should develop such a description coupling the suspension cell and flow dynamics in response to gravity, flow and light, as has been done by Williams and Bees to describe bioconvection patterns [9]. Numerical and analytical predictions from such models will predict the spatio-temporal patterns in the suspension, including the formation of propagating waves of cells and their concentration in the phototactic curtain structures we have observed. To describe the latter, accounting for the observed width of the curtain pattern, it will be necessary to develop a model coupling the local light profile (optical shadows from the mesh) to the photogyrotactic dynamics. Photogyrotactic models should be developed for the lower, porous and upper regions combined, and should be able predict the characteristic timescales we have observed, such as the time required for plumes to form off the upper surface (SM Figure 3). Such models will also describe how the plume-cloud in the harvest region grows with time, accounting for the curvature in the meniscus (neglected in our essential model) and how this affects the plume formation. Observation indicates that the plume forms in the lowest point of the meniscus, likely because cells accumulate there. Advanced modelling should also predict how long the mesh or porous layer is able to support the plume against sinking when the LED is on, and how long it takes to sink through the layer when the LED is switched off.

A full account of photogyrotactic dynamics will permit inclusion of processes (such as diffusion and light shading affecting phototactic speed) not included in our essential model. Predictions from these improved photogyrotactic models for the concentration in the upper, porous and lower regions, should provide better agreement with the results shown in Figure 6. In particular, it will be interesting to use these refined models to establish the parameters that determine optimal conditions for harvesting microalgae in the upper region. From a practical perspective it is desirable to obtain the largest possible volume of suspension with the highest concentration gain for a given initial mean concentration and critical parameters, such as the height of the lower, porous and upper harvest regions, and the total duration of the concentration process. In addition, it will be desirable to know how strong the light intensity should be for optimal phototactic concentration. This is a parameter which was held fixed in the present study.

Following a demonstration that photogyrotactic concentration can be effective at the milliliter ('cuvette') scale, future investigations should consider how this new concentration method can be scaled up to be of value in industrial microalgal bioprocessing and harvesting. As mentioned in the introduction, harvesting contributes a significant amount (about 20-30% [32]) of the cost for processing microalgae and bioproducts derived from them. For industrially-valuable swimming microalgae, exploiting swimming in response to light, as we have here explored, has not been considered as the basis for an efficient new harvesting method. *Dunaliella salina*, a marine relative of *C. reinhardtii*, is cultured in ponds that are maximum 20 cm deep to allow light penetration for growth [39]. It is known that this microalga can be concentrated when a layer of freshwater is produced, artificially or by rain, at the surface of the pond [39]. The freshwater generates a gradient in the density of the suspension medium, which acts similarly to the porous layer in our study and causes the microalgae to become trapped in the freshwater layer at the surface [40]. The role of photogyrotaxis in this industrially well-known concentration process [39] has not yet been investigated. However, taking into consideration the concentration physics we have uncovered in this study, it could be optimized to produce better microalgal yields from culture ponds. Density gradients cannot be exploited for freshwater microalgae (an aqueous suspending medium less dense than water is not easily found), which require a porous layer to be concentrated by upswimming. In this case, the use of glass beads for the porous layer, as in this study, represents an improvement over Kessler's original suggestion of a fibrous porous layer [31], which, from experience with gravitactic concentration using cotton wool [14,15,41], is known to be liable to irreversible cell loss to the fibers (biofouling).

Finally, it is worth remembering that *C. reinhardtii* is a soil-dwelling microalga. Little is known about its ecology within soils [42], but we can speculate that in saturated soils *C. reinhardtii* may migrate across porous layers in response to daylight. Thus, the phenomenology we have uncovered in this work and the methods we have developed can be adapted to better understand the behaviour of *C. reinhardtii* and similar species in their natural environments. It will be very interesting in future studies to investigate the phototactic movements of *C. reinhardtii* in laboratory soil-like porous media, and how this social behaviour affects its photosynthetic growth in topsoil, as well as more 'traditional' social behaviours, such as sex [42] and interactions with other soil microbes [43,44].

## 5    Data availability

Data and MATLAB code for this study are available from the Zenodo repository: https://doi.org/10.5281/zenodo.5113916.

## 6    Author Contributions

PP carried out the experiments. OAC developed the modelling. PP and OAC designed the experiments, analysed the data, and wrote the paper.

## 7 Acknowledgments

We thank K. Leptos for providing the *C. reinhardtii* CC125, M. Bees for a critical reading of the manuscript, and H. Laeverenz Schlogelhofer for the laboratory training provided to PP. OAC acknowledges financial support from the Winton Programme for the Physics of Sustainability. PP and OAC acknowledge financial support from the British Council through the Newton Bhabha Fellowship Scheme.

## Appendix A

Derivation of the Essential Model from the full photogyrotactic Williams and Bees Model

We show here how our essential model can be derived from the full photogyrotactic model of Williams and Bees [9], using the assumptions we have listed in the main text. We will first briefly summarise the latter model. The model considers flow in a suspension, described by the Navier-Stokes equation:

$$\rho\left(\frac{\partial \boldsymbol{u}}{\partial t} + (\boldsymbol{u} \cdot \nabla)\boldsymbol{u}\right) = -\nabla P + \nabla \cdot \boldsymbol{\Sigma} + c\, v_c\, \Delta\varrho\, \mathbf{g}, \tag{A1}$$

where $\rho$ is the suspending fluid density, $\boldsymbol{u}$ is the flow speed, $P$ is the pressure, $\boldsymbol{\Sigma}$ is the net stress experienced by the suspension (comprising passive and active viscous stresses). The last term quantifies the source of flow driven by the buoyancy of cells with excess density $\Delta\varrho$, volume $v_c$ and cell concentration $c$; $\mathbf{g}$ is the acceleration due to gravity. The flow is assumed incompressible so that $\nabla \cdot \boldsymbol{u} = 0$. We then have a continuity equation for the cell density $c$, expressing cell conservation:

$$\frac{\partial c}{\partial t} = -\nabla \cdot \left[\, (\boldsymbol{u} + V(I)\langle\boldsymbol{p}\rangle(I))c - \mathbf{D}(I) \cdot \nabla c \,\right]. \tag{A2}$$

where $V(I)$ is the light intensity-dependent swimming speed of the cells, and $\langle\boldsymbol{p}\rangle(I)$ and $\mathbf{D}(I)$ are the light intensity-dependent mean orientation and diffusivity tensor, respectively. Bees and Williams chose the speed to be a linear function of the light intensity (photokinesis), while the mean orientation and diffusivity were derived from a steady state Fokker-Plank equation:

$$\nabla_p \cdot \left[\, \dot{\boldsymbol{p}}\, f - \mathrm{D}_R\, \nabla_p f \,\right] = 0 \tag{A3}$$

where subscript $p$ denotes derivatives with respect to orientation, $\mathrm{D}_R$ is the rotational diffusivity of a cell, and $f$ is the probability density that a cell will have orientation $\boldsymbol{p}$, with dynamics governed by:

$$\dot{\boldsymbol{p}} = \alpha_g(I)[\boldsymbol{k} - (\boldsymbol{k} \cdot \boldsymbol{p})\boldsymbol{p}] + \alpha_p(I)[\boldsymbol{p} \times (\beta_1\, \boldsymbol{\pi} + \beta_2\, \nabla I)] \times \boldsymbol{p} - \frac{1}{2}\,\boldsymbol{\omega} \times \boldsymbol{p}, \tag{A4}$$

where the first term in the equation derives from the torque due to gravity. The reorientation frequency $\alpha_g(I)$ is in general assumed to depend on light intensity (light could change the degree of bottom-heaviness of a cell, changing the gravitational torque). The second term comes from a phenomenological torque due to phototaxis, with cells either responding to light in a certain direction $\boldsymbol{\pi}$ or to a light gradient $\nabla I$, or a linear combination of the two governed by the coefficients $\beta_1$ and $\beta_2$. The third term corresponds to reorientation from the flow vorticity $\boldsymbol{\omega}$, where we have assumed for simplicity that cells are spherical (see Bees and Williams [9] for the equations for ellipsoidal cells). Once the probability density is obtained from solving equations (A3) and (A4), the mean orientation is given by $\langle \boldsymbol{p} \rangle = \int_{\boldsymbol{p}} \boldsymbol{p} \, f(\boldsymbol{p}) \, d\boldsymbol{p}$, where we have integrated over orienations $\boldsymbol{p}$. Following Pedley and Kessler [2], the diffusivity tensor is approximated as $\boldsymbol{D} = D \langle (\boldsymbol{p} - \langle \boldsymbol{p} \rangle)(\boldsymbol{p} - \langle \boldsymbol{p} \rangle) \rangle$, where $D$ is a characteristic diffusivity scale. This is a good approximation for weak flows [15]. Finally, the effect of light shading was modelled by using the Beer-Lambert law to evaluate the intensity I at depth z, $I(z) = I_s \, exp \left\{ -\varepsilon \int_z^0 n(z) \, dz \right\}$, where $I_s$ is the intensity at the source (z = 0) and $\varepsilon$ is the extinction coefficient for a cell.

Next, we apply our assumptions to derive the essential model. In particular, for cell reorientation by real and effective torques we assume that the phototactic reorientation is dominant (assumption vi in section 2.2 of the main text), so that in equation (A4) $\alpha_p \ll \alpha_g$, $|\boldsymbol{\omega}|$, where following [24] $\alpha_p$ is assumed independent of the light intensity $I$. We further assume that the microalgae reorient only to the direction of light (model C, case I in [9], our assumption ii), in our case straight upwards along a vertically oriented unit vector $\boldsymbol{k}$, and not to its gradient. Thus, setting $\beta_1 = 1$ and $\beta_2 = 0$, equation (A4) becomes

$$\dot{\boldsymbol{p}} = \alpha_p [\boldsymbol{p} \times \boldsymbol{k}] \times \boldsymbol{p} = \alpha_p [\boldsymbol{k} - (\boldsymbol{k} \cdot \boldsymbol{p}) \boldsymbol{p}], \qquad (A5)$$

where in the last step we have used a standard vector identity to show that, in our simplified model, the phototactic reorientation has the same form as the gravitactic reorientation (compare with the first term in equation (A4)). Substituting (A5) into equation (A3) and non dimensionalising by dividing through by the rotational diffusivity $D_R$, we obtain

$$\nabla_p \cdot \left[ \lambda_p [\boldsymbol{k} - (\boldsymbol{k} \cdot \boldsymbol{p}) \boldsymbol{p}] f - \nabla_p f \right] = 0, \qquad (A6)$$

where $\lambda_p = \frac{\alpha_p}{2 D_R}$ is a nondimensional ratio of reorientation by phototaxis to that by rotational diffusion. The solution of equation (A6) has been previously obtained for gyrotactic suspensions [2], and is given by a Von Mises distribution function

$$f(\theta) = \mu e^{\lambda_p \, cos(\theta)} \qquad (A7)$$

where $\mu = \lambda_p / 4\pi \sinh(\lambda_p)$ and $\theta$ is the polar angle the vertical. The mean swimming direction, which enters into the continuity equation (A2), is then given by

$$\langle \boldsymbol{p} \rangle = \int_S \boldsymbol{p} f(\boldsymbol{p}) \, dS = \int_0^{2\pi} \int_0^{\pi} \boldsymbol{p} \, f(\theta) \, sin(\theta) \, d\theta d\phi = (0, 0, P_0) \qquad (A8)$$

where $P_0 = \coth(\lambda_p) - 1/\lambda_p$. So, as might be expected, cells swim upwards on average. The diffusivity can be similarly computed from the expression given further above. As shown in [2], it is given by the diagonal matrix $\boldsymbol{D} = D \, diag(D_h, D_h, D_v)$, where $D$ is the diffusivity scale (see Table 1), and where the horizontal diffusivity components are $D_h = P_0/\lambda_p$ and the vertical is $D_v = 1 - \coth(\lambda_p)^2 + 1/\lambda_p^2$. We notice that since $\alpha_p$, and thus $\lambda_p$, have been assumed not to depend on the light intensity $I$, then neither do $\langle \boldsymbol{p} \rangle$ and $\boldsymbol{D}$, contrary to what was assumed in general by the Williams and Bees model [9]. We next assume that phototactic reorientation dominates rotational diffusion (true for moderately large $\lambda_p$), so that $P_0 = 1$, and that diffusion is isotropic, $D_h/D_v \approx 1$ (assumption vii). This will never be strictly true, but for moderate values of $\lambda_p$ it is an acceptable approximation, which has also been made in other models of *C. reinhardtii* swimming in shear flows [45] and undergoing phototaxis [24,26]. Then in our essential model we have $\langle \boldsymbol{p} \rangle \approx (0,0,1) = \boldsymbol{k}$ and $\boldsymbol{D} = D\,\boldsymbol{I}$, where $\boldsymbol{I}$ denotes the identity matrix. With these simplifications, the continuity equation (A2) for the microalgae, assuming also that the swimming speed does not depend on intensity and is equal to the mean swimming speed $V(I) \approx V_s$ (assumption vi), and that background flows are negligible compared to swimming $|\boldsymbol{u}| \ll V_s$ (assumption v), reduces to equation (6) in the main text.

Appendix B

Essential model for the mesh-only case

Summarised here is the derivation for the essential model when there is only a mesh dividing upper and lower regions. As discussed in the methods and main text, the mesh pores are large compared to the size of the microalgae, so it is assumed that swimmers can pass them freely, with no effect on the swimming speed. In this case, the balance equations for the swimmer number in the upper and lower regions are given by:

$$\frac{d\overline{N}_l^m}{dt} = -A\,V_s\,\bar{c}_l^m, \tag{B1a}$$

$$\frac{d\overline{N}_u^m}{dt} = A\,V_s\,\bar{c}_l^m. \tag{B1b}$$

where all quantities are as in the main text, the superscript '$m$' denoting cell numbers and concentrations in the mesh-only case. We see that in this case, there is a flux out of the lower region, of concentration $\bar{c}_l$, due to cells swimming upwards into the upper region with speed $V_s$. Correspondingly, an equal and opposite flux enters the upper region. Recalling $\overline{N}_i^m = Ah_i\,\bar{c}_i^m$, and substituting into (B1), we obtain, dividing both sides of the equations by the respective $h_i$,

$$\frac{d\bar{c}_l^m}{dt} = -\alpha\,\bar{c}_l^m, \tag{B2a}$$

$$\frac{d\bar{c}_u^m}{dt} = -\eta\,\bar{c}_l^m. \tag{B2b}$$

where, as in the main text, $\alpha = V_s/h_l$ and $\eta = V_s/h_u$. The first equation the same as for the porous layer model. We can substitute its solution into (B2b) to find that the concentration in the upper and lower regions is given by equations (4) in the main text. Using the parameters of Table 2 below, these equations were plotted using MATLAB (The MathWorks, Natick, MA) to obtain the profiles in Figure 7a (

**Table 2. Essential model parameters for mesh-only case.** Values were obtained from direct measurements of our experimental system. All swimmer parameters for this model are the same as in Table 1.

| Parameter | Symbol | Units | Axenic algae | Reference |
|---|---|---|---|---|
| Lower region height | $h_l$ | $cm$ | 2.14 | This work |
| Porous region height | $h_p$ | $cm$ | 0.41 | This work |
| Initial mean concentration of suspension in the lower region | $\bar{c}_l^{m0}$ | $cells\ cm^{-3}$ | $1.20 \times 10^6$ | This work |
| Initial mean concentration of suspension in the upper region | $\bar{c}_u^{m0}$ | $cells\ cm^{-3}$ | $1.12 \times 10^6$ | This work |
| Upswimming rate 1 | $\alpha = \dfrac{V_s}{h_l}$ | $s$ | $3.70 \times 10^{-3}$ | This work |
| Upswimming rate 2 | $\eta = \dfrac{V_s}{h_u}$ | $s$ | $1.98 \times 10^{-2}$ | This work |

## References

1. Wadhams GH, Armitage JP. Making sense of it all: bacterial chemotaxis. *Nat Rev Mol Cell Biol* (2004) **5**:1024-1037. doi:10.1038/nrm1524

2. Pedley TJ, Kessler JO. Hydrodynamic Phenomena in Suspensions of Swimming Microorganisms. *Annu Rev Fluid Mech* (1992) **24**:313–358. doi:10.1146/annurev.fl.24.010192.001525

3. Foster KW, Smyth RD. Light Antennas in phototactic algae. *Microbiol Rev* (1980) **44**: 572-630. doi:10.1128/mr.44.4.572-630.1980

4. Adler J. Chemotaxis in bacteria. *Science* (1966) **153**:708–716. doi:10.1126/science.153.3737.708

5. Bhattacharjee T, Amchin DB, Alert R, Ott JA, Datta SS. Chemotactic smoothing of collective migration. (2021). http://arxiv.org/abs/2101.04576

6.    Bretschneider T, Othmer HG, Weijer CJ. Progress and perspectives in signal transduction, actin dynamics, and movement at the cell and tissue level: lessons from *Dictyostelium*. *Interface Focus* (2016) **6**:20160047. doi:10.1098/rsfs.2016.0047

7.    Bees MA. Advances in Bioconvection. *Annu Rev Fluid Mech* (2020) **52**:449–476. doi:10.1146/annurev-fluid-010518-040558

8.    O'Malley S, Bees MA. The Orientation of Swimming Biflagellates in Shear Flows. *Bull Math Biol* (2012) **74**:232–255. doi:10.1007/s11538-011-9673-1

9.    Williams CR, Bees MA. Photo-gyrotactic bioconvection. *J Fluid Mech* (2011) **678**:41–86. doi:10.1017/jfm.2011.100

10.   Bees MA, Croze OA. Dispersion of biased swimming micro-organisms in a fluid flowing through a tube. *Proc R Soc A Math Phys Eng Sci* (2010) **466**:2057–2077. doi:10.1098/rspa.2009.0606

11.   Bearon RN, Bees MA, Croze OA. Biased swimming cells do not disperse in pipes as tracers: A population model based on microscale behaviour. *Phys Fluids* (2012) **24**:121902. doi:10.1063/1.4772189

12.   Hwang Y, Pedley TJ. Bioconvection under uniform shear: linear stability analysis. *J Fluid Mech* (2014) **738**:522–562. doi:10.1017/jfm.2013.604

13.   Cencini M, Franchino M, Santamaria F, Boffetta G. Centripetal focusing of gyrotactic phytoplankton. *J Theor Biol* (2016) **399**:62-70. doi:10.1016/j.jtbi.2016.03.037

14.   Croze OA, Ashraf EE, Bees MA. Sheared bioconvection in a horizontal tube. *Phys Biol* (2010) **7**:46001. doi:10.1088/1478-3975/7/4/046001

15.   Croze OA, Bearon RN, Bees MA. Gyrotactic swimmer dispersion in pipe flow: Testing the theory. *J Fluid Mech* (2017) **816**:481-506. doi:10.1017/jfm.2017.90

16.   Durham WM, Kessler JO, Stocker R. Disruption of Vertical Motility by Shear Triggers Formation of Thin Phytoplankton Layers. *Science* (2009) **323**:1067–1071. doi: 10.1126/science.1167334

17.   Elgeti J, Winkler RG, Gompper G. Physics of microswimmers—single particle motion and collective behavior: a review. *Reports Prog Phys* (2015) **78**:056601 doi:10.1088/0034-4885/78/5/056601

18.   Rusconi R, Stocker R. Microbes in flow. *Curr Opin Microbiol* (2015) **25**:1–8. doi:10.1016/j.mib.2015.03.003

19.   Marchetti MC, Joanny JF, Ramaswamy S, Liverpool TB, Prost J, Rao M, Simha RA. Hydrodynamics of soft active matter. *Rev Mod Phys* (2013) **85**:1143–1189. doi:10.1103/RevModPhys.85.1143

20.   Elgeti J, Winkler RG, Gompper G. Physics of microswimmers—single particle motion and collective behavior: a review. *Rep Prog Phys* (2015) **78**:56601.

21.  Bastos-Arrieta J, Revilla-Guarinos A, Uspal WE, Simmchen J. Bacterial biohybrid microswimmers. *Front Robot AI* (2018) **5**: doi:10.3389/frobt.2018.00097

22.  Prakash P, Abdulla AZ, Singh V, Varma M. Tuning the torque-speed characteristics of the bacterial flagellar motor to enhance swimming speed. *Phys Rev E* (2019) **100**:062609. doi:10.1103/PhysRevE.100.062609

23.  Prakash P, Abdulla AZ, Singh V, Varma M. Swimming statistics of cargo-loaded single bacteria. *Soft Matter* (2020) **16**:9499–9505. doi:10.1039/d0sm01066a

24.  Arrieta J, Barreira A, Chioccioli M, Polin M, Tuval I. Phototaxis beyond turning: Persistent accumulation and response acclimation of the microalga Chlamydomonas reinhardtii. *Sci Rep* (2017) **7**:1–7. doi:10.1038/s41598-017-03618-8

25.  Leptos KC, Chioccioli M, Furlan S, Pesci AI, Goldstein RE. An Adaptive Flagellar Photoresponse Determines the Dynamics of Accurate Phototactic Steering in Chlamydomonas. *bioRxiv* (2018)254714. doi:10.1101/254714

26.  Arrieta J, Polin M, Saleta-Piersanti R, Tuval I. Light Control of Localized Photobioconvection. *Phys Rev Lett* (2019) **123**:158101. doi:10.1103/PhysRevLett.123.158101

27.  Ogawa T, Shoji E, Suematsu NJ, Nishimori H, Izumi S, Awazu A, Iima M. The Flux of Euglena gracilis Cells Depends on the Gradient of Light Intensity. *PLoS One* (2016) **11**:e0168114. doi:10.1371/journal.pone.0168114

28.  Dervaux J, Capellazzi Resta M, Brunet P. Light-controlled flows in active fluids. *Nat Phys* (2017) **13**:306–312. doi:10.1038/nphys3926

29.  Javadi A, Arrieta J, Tuval I, Polin M. Photo-bioconvection: Towards light control of flows in active suspensions: Photo-bioconvection. *Philos Trans R Soc A Math Phys Eng Sci* (2020) **378**: doi:10.1098/rsta.2019.0523

30.  Williams CR, Bees MA. A tale of three taxes: Photo-gyro-gravitactic bioconvection. *J Exp Biol* (2011) **214**:2398–2408. doi:10.1242/jeb.051094

31.  US4324067A - Algal cell harvesting https://patents.google.com/patent/US4324067A/en?oq=U.S.+Patent+No.+4324.067

32.  Fasaei F, Bitter JH, Slegers PM, van Boxtel AJB. Techno-economic evaluation of microalgae harvesting and dewatering systems. *Algal Res* (2018) **31**:347–362. doi:10.1016/j.algal.2017.11.038

33.  Gorman DS, Levine RP. Cytochrome f and plastocyanin: their sequence in the photosynthetic electron transport chain of Chlamydomonas reinhardi. *Proc Natl Acad Sci U S A* (1965) **54**:1665–9. doi:10.1073/pnas.54.6.1665

34.  Jin D, Kotar J, Silvester E, Leptos KC, Croze OA. Diurnal Variations in the Motility of Populations of Biflagellate Microalgae. *Biophys J* (2020) **119**:2055–2062. doi:10.1016/j.bpj.2020.10.006

35. Jakuszeit T, Croze OA, Bell S. Diffusion of active particles in a complex environment: Role of surface scattering. *Phys Rev E* (2019) **99**: doi:10.1103/PhysRevE.99.012610

36. Théry A, Wang Y, Dvoriashyna M, Eloy C, Elias F, Lauga E. Rebound and scattering of motile *Chlamydomonas* algae in confined chambers. *Soft Matter* (2021) **17**:4857–4873. doi:10.1039/D0SM02207A

37. Goldstein RE, Polin M, Tuval I. Emergence of synchronized beating during the regrowth of eukaryotic flagella. *Phys Rev Lett* (2011) **107**:148103. doi:10.1103/PhysRevLett.107.148103

38. Kessler JO. Hydrodynamic focusing of motile algal cells. *Nature* (1985) **313**:218–220. doi:10.1038/313218a0

39. Borowitzka LJ, Borowitzka MA. Commercial production of β-carotene by Dunaliella salina in open ponds. *Bull Mar Sci* (1990) **47**:244–252.

40. Bearon RN, Grünbaum D. Bioconvection in a stratified environment: Experiments and theory. *Phys Fluids* (2006) **18**:127102. doi:10.1063/1.2402490

41. Kessler JO. "The external dynamics of swimming micro-organisms," in *Progress in Phycological Research* (Bristol: Biopress), 257–307.

42. Sasso S, Stibor H, Mittag M, Grossman AR. The Natural History of Model Organisms: From molecular manipulation of domesticated Chlamydomonas reinhardtii to survival in nature. *Elife* (2018) **7**:e39233. doi:10.7554/eLife.39233

43. Teplitski M, Rajamani S. Signal and Nutrient Exchange in the Interactions Between Soil Algae and Bacteria. *Soil Biology* (2010) **23:**413-426. doi:10.1007/978-3-642-14512-4_16

44. Schlogelhofer HL, Peaudecerf FJ, Bunbury F, Whitehouse MJ, Foster RA, Smith AG, Croze OA. Combining SIMS and mechanistic modelling to reveal nutrient kinetics in an algal-bacterial mutualism. *PLoS One* (2021) **16**:e0251643. doi:10.1371/journal.pone.0251643

45. Barry MT, Rusconi R, Guasto JS, Stocker R. Shear-induced orientational dynamics and spatial heterogeneity in suspensions of motile phytoplankton. *J R Soc Interface* (2015) **12**:20150791. doi: 10.1098/rsif.2015.0791

# *Supplementary Material*

## Phototgyrotactic concentration of a population of swimming microalgae across a porous layer

Praneet Prakash[1], Ottavio A. Croze [2,*]

[1] Department of Applied Mathematics and Theoretical Physics, Centre for Mathematical Sciences, University of Cambridge, Cambridge CB3 0WA, United Kingdom

[2] School of Mathematics, Statistics and Physics, Newcastle University, Newcastle upon Tyne NE1 7RU, United Kingdom

*Corresponding Author
otti.croze@newcastle.ac.uk

## 1.     Constituents of Growth Media

The Tris-minimal growth media is prepared by mixing following components (A, B, C, D) and then topped up by deionized (DI) water to reach a final volume of 1 L.

A) 20 mL of 1 M Tris Solution

B)  25 mL of Salt Solution.

The Salt Solution 'B' is prepared by adding the following three salts in 1 L DI water:

   1)  $NH_4Cl$ – 15.0 gm
   2)  $MgSO_4 \cdot 7H_2O$ – 4.0 gm
   3)  $CaCl_2 \cdot 2H_2O$ – 2.0 gm

C) 1 mL of Phosphate Solution

The Phosphate Solution 'C' is prepared by adding the following two phosphates in 100 mL DI water:

   1)  $K_2HPO_4$ – 10.8 gm
   2)  $KH_2PO_4$ – 5.6 gm

D) 1 mL each of 7 Trace Elements

The table below details out the preparation of Trace Elements:

| Number | Chemical Component | Concentration | Mass per 50 mL stock |
|--------|--------------------|---------------|----------------------|
| 1 | $EDTA \cdot Na_2 \cdot 2H_2O$ | 25 mM | 0.465 gm |
| 2 | $(NH_4)_6Mo_7O_{24} \cdot 4H_2O$ | 32 $\mu$m | 0.002 gm |
| 3 | $CuCl_2 \cdot 2H_2O$ | 1.4 mM | 0.017 gm |
|   | EDTA | 2 mM | 0.029 gm |

| 4 | ZnSo$_4 \cdot$7H$_2$O | 2.5 mM | 0.036 gm |
|---|---|---|---|
|   | EDTA | 2.7 mM | 0.040 gm |
| 5 | MnCl$_2 \cdot$4H$_2$O | 6 mM | 0.059 gm |
|   | EDTA | 6 mM | 0.088 gm |
| 6 | FeCl$_3 \cdot$6H$_2$O | 20 mM | 0.270 gm |
|   | EDTA | 22 mM | 0.321 gm |
|   | Na$_2$CO$_3$ | 22 mM | 0.116 gm |
| 7 | CoCl$_2 \cdot$6H$_2$O | 7 mM | 0.083 gm |

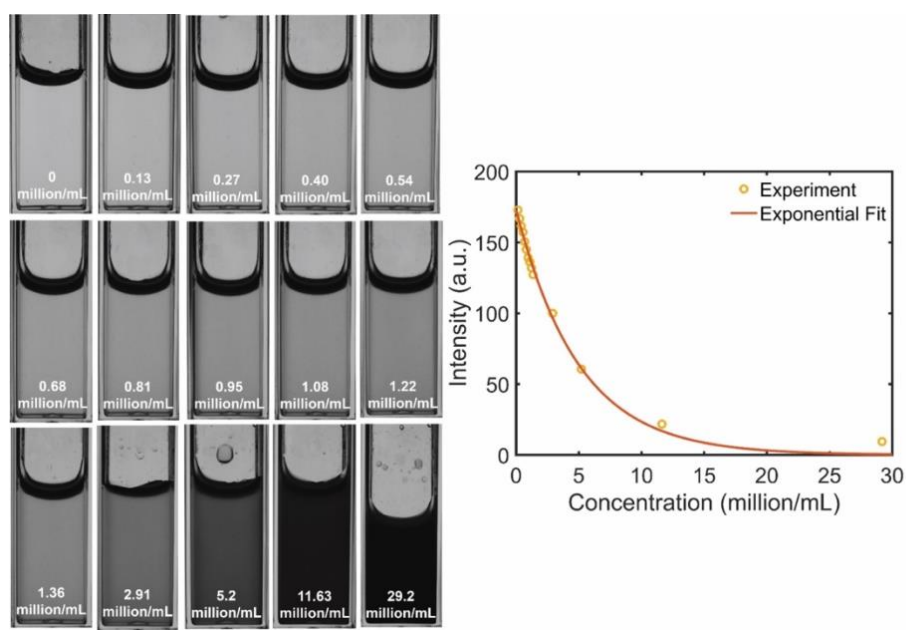

**SM Figure 1.** Calibration curve of light intensity versus concentration of algae.

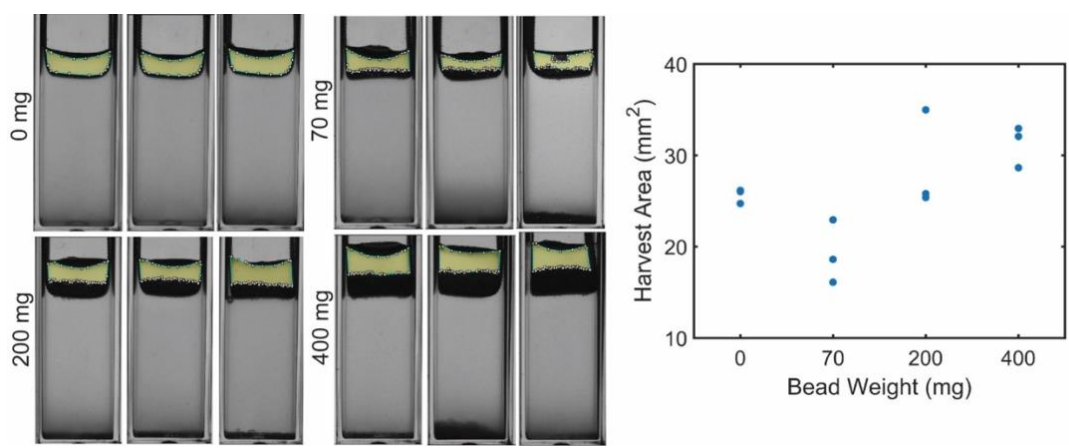

**SM Figure 2.** The yellow shaded area is the harvest region used to determine the integrated intensity of the microalgae. Figure on the right shows harvest area used for estimating intensity and thereby average concentration in each case.

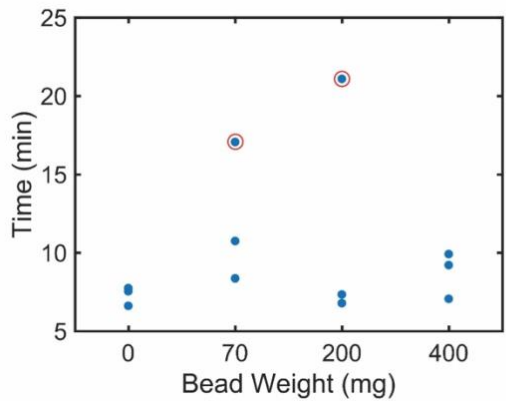

**SM Figure 3.** Average time taken to form the plume as a function of bead layer weight (0 mg corresponds to the mesh-only case). The time is between $6 - 9$ min in nearly all the cases with an average of 8.1 min excluding the two outliers corresponding to atypical samples (marked by red rings) beyond 15 min.

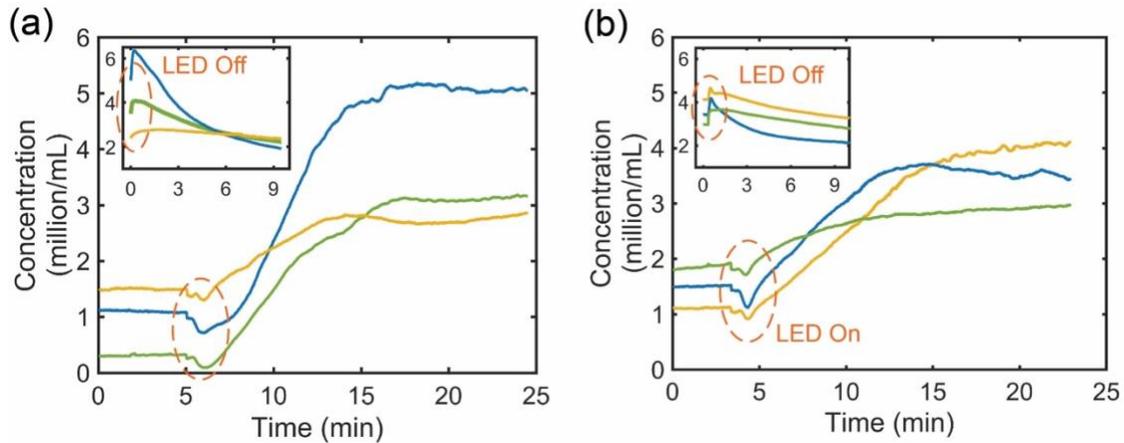

**SM Figure 4.** Trial temporal concentration profiles for the porous layers weighing 70 mg (a) and 200 mg (b), after the LED is switched on and off (inset), as shown. These cases were found to be variable (some repeats were more leaky than others).